\documentclass[10pt,conference,anonymous]{IEEEtran}
\IEEEoverridecommandlockouts
\usepackage{array}
\usepackage{amsmath,amsfonts}
\usepackage{algorithmic}
\usepackage{graphicx}
\usepackage{textcomp}
\usepackage{xcolor}
\usepackage{tcolorbox}
\usepackage{listings}
\usepackage{makecell}
\usepackage{adjustbox}
\usepackage{multirow}
\usepackage{svg}
\usepackage{xspace}
\usepackage{amsmath}
\usepackage{subfigure}
\usepackage{amsmath}
\usepackage{booktabs}
\usepackage{color, colortbl}
\usepackage{url}
\usepackage{graphicx}
\usepackage{caption} 
\usepackage{booktabs}
\usepackage{pifont}
\usepackage{dsfont}
\usepackage{floatrow}
\newfloatcommand{capbtabbox}{table}[][\FBwidth]
\usepackage{tabularx}

\DeclareUnicodeCharacter{2212}{-}
\definecolor{Gray}{gray}{0.9}
\definecolor{green}{RGB}{102,252,102}
\definecolor{ored}{RGB}{255,99,71}
\definecolor{orange}{RGB}{255,165,0}
\definecolor{lightgray}{RGB}{211,211,211}

\usepackage[ruled,vlined,linesnumbered]{algorithm2e}

\SetCommentSty{mycommfont}

\usepackage{tikz}
\usetikzlibrary{shapes}

\newcommand{\tool}{{\texttt{CAPdroid}}\xspace}

\newcommand{\chen}[1]{\textcolor{red}{\textbf{Chen}: #1}}

\makeatletter
\newcommand{\linebreakand}{%
  \end{@IEEEauthorhalign}
  \hfill\mbox{}\par
  \mbox{}\hfill\begin{@IEEEauthorhalign}
}
\makeatother

\definecolor{amber}{rgb}{1.0, 0.49, 0.0}

\def\BibTeX{{\rm B\kern-.05em{\sc i\kern-.025em b}\kern-.08em
    T\kern-.1667em\lower.7ex\hbox{E}\kern-.125emX}}
\begin{document}

\title{Read It, Don't Watch It: Captioning Bug Recordings Automatically}

\author{\IEEEauthorblockN{Sidong Feng\IEEEauthorrefmark{2}, Mulong Xie\IEEEauthorrefmark{3}, Yinxing Xue\IEEEauthorrefmark{4}, Chunyang Chen\IEEEauthorrefmark{2}\IEEEauthorrefmark{1}}
\IEEEauthorblockA{\IEEEauthorrefmark{2} Monash University \\
\IEEEauthorrefmark{3} Australian National University \\
\IEEEauthorrefmark{4} University of Science and Technology of China \\ 
Email: \IEEEauthorrefmark{2}\{sidong.feng,chunyang.chen\}@monash.edu,
\IEEEauthorrefmark{3}mulong.xie@anu.edu.au,
\IEEEauthorrefmark{4}yxxue@ustc.edu.cn} \\
\thanks{\IEEEauthorrefmark{1} Corresponding author}}


\maketitle

\begin{abstract}
Screen recordings of mobile applications are easy to capture and include a wealth of information, making them a popular mechanism for users to inform developers of the problems encountered in the bug reports. However, watching the bug recordings and efficiently understanding the semantics of user actions can be time-consuming and tedious for developers. Inspired by the conception of the video subtitle in movie industry, we present a lightweight approach \tool to caption bug recordings automatically. \tool is a purely image-based and non-intrusive approach by using image processing and convolutional deep learning models to segment bug recordings, infer user action attributes, and generate subtitle descriptions. The automated experiments demonstrate the good performance of \tool in inferring user actions from the recordings, and a user study confirms the usefulness of our generated step descriptions in assisting developers with bug replay.
\end{abstract}


\section{Introduction}
Software maintenance activities are known to be generally expensive, and challenging~\cite{planning2002economic} and one of the most important maintenance tasks is to handle bug reports~\cite{anvik2005coping}.
A good bug report is detailed with clear information about what happened and the steps to reproduce the bug.
However, writing such clear and concise bug reports takes time, especially for non-developer or non-tester users who do not have that expertise and are not willing to spend that much effort~\cite{aranda2009secret, feng2022gifdroid}.
The emergence of screen recording significantly lowers the bar for bug documenting. 
First, it is easy to record the screen as there are many tools available, some of which are even embedded in the operating system by default, like iOS~\cite{web:iosrecord} and Android~\cite{web:androidrecord}.
Second, video recording can include more detail and context such as configurations, and parameters, bridging the understanding gap between users and developers.   

\begin{figure}
	\centering
	\subfigure[Default]{
		\includegraphics[width = 0.21\linewidth]{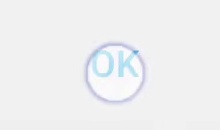}
		\label{fig:default}}
	\hfill
	\subfigure[Cursor]{
		\includegraphics[width = 0.21\linewidth]{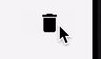}
		\label{fig:cursor}}	
	\hfill
	\subfigure[Custom]{
		\includegraphics[width = 0.21\linewidth]{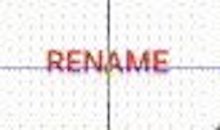}
		\includegraphics[width = 0.21\linewidth]{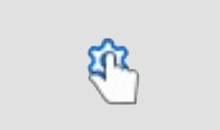}
		\label{fig:custom}}	
	\caption{Examples of touch indicators.}
	\label{fig:touch}
\end{figure}

Unfortunately, in many cases, watching the bug recordings and understanding the user behaviors can be time-consuming and tedious for developers~\cite{feng2021gifdroid,bernal2020translating}.
First, the recording may play too fast to watch, and the developers have to pause the recording, or even replay it multiple times to recognize the bug.
Second, the watching experience can be further deteriorated by blurred video resolution, poor video quality, etc.
Third, the recording usually contains a visual indicator (in Fig.~\ref{fig:touch}) to help developers identify the user actions performed on the screen.
However, those indicators sometimes are too small to be conspicuously realized, and developers have to the recording back and forth to guess each action to repeat it in their testing environment.

Besides bug recordings, those issues also apply to general videos (e.g., movies, drama, etc).
To address those issues in normal video watching, captions or subtitles are provided to add clarity of details, better engage users, maintain concentration for longer periods, and translate the different languages~\cite{gernsbacher2015video,garza1991evaluating}.
Inspired by the conception of video subtitles in the movie industry, we intend to generate the caption of an app recording to add analogous benefits to developers. 
Given a caption accompanying the recording, developers, especially novices can more easily identify the user behaviors in the recording and shift their focus toward bug fixing.
Specifically, we segment recordings into clips to characterize the “scenes” in the movie and add action descriptions of each clip to guide developers.

Existing work has investigated methods to generate a textual description for GUI screenshot~\cite{chen2020unblind, moran2022empirical,chen2018ui,feng2021auto,feng2022auto}, which has been shown useful for various downstream tasks such as GUI retrieval, accessibility enhancement, code indexing, etc.
Chen et al.~\cite{chen2020unblind} propose an image captioning model to apply semantic labels to GUI elements to improve the accessibility of mobile apps.
Clarity~\cite{moran2022empirical} further consider multi-modal GUI sources to generate high-level descriptions for the entire GUI screen.
However, none of them can generate descriptions for video recording, which is a more challenging task, translating spatial and temporal information into a semantic natural language.


To create good video subtitles, there are several standards~\cite{cintas2014audiovisual}, including caption synchronization with the videos, accurate content comprehension, compact and consistent word usage, etc.
Similarity, we propose an image-based approach \tool in this paper to non-intrusively caption each action step for a bug recording, including three phases: 1) \textit{action segmentation}, 2) \textit{action attribute inference}, and 3) \textit{description generation}.
Inspired by the previous work GIFdroid~\cite{feng2022gifdroid,feng2022gifdroid2} to localize keyframes in bug recording, we first develop a heuristic method to segment the recording into a sequence of action clips (i.e., TAP, SCROLL, INPUT).
Then, we adopt image-processing and deep-learning methods to model the spatial and temporal features across frames in the clips to infer action attributes, such as touch location, moving offset, and input text.
A simple description based on the action attribute, e.g. tap on (x,y) coordinate, cannot express the action intuitively.
Therefore, we first utilize off-the-shelf GUI models to non-intrusively gather the elements information in the GUI.
As the GUI elements of interest may not have enough context to be uniquely identified, we propose a novel algorithm using global information of GUI elements to generate high-level semantic descriptions.

We first evaluate the performance of the \tool in obtaining user actions by \textit{action segmentation} and \textit{action attribute inference}, through an automated method.
We collect 439 Android apps from Google Play and leverage an automated app explore tool to simulate user actions on the screen, meanwhile capturing a 10-min screen recording for each app.
Results show that our tool achieves the best performance (0.84 Video F1-score and 0.93 accuracy) in action segmentation from the recordings compared with five commonly-used baselines. 
\tool also achieves on average 91.46\% in inferring action attributes, outperforming two state-of-the-art baselines.
We further carry out a user study to evaluate the usefulness of \textit{description generation} of \tool in assisting bug replay, with 10 real-world bug recordings from GitHub.
Results show that participants save 59.8\% time reproducing the bug with the help of the steps we described, compared with the steps written by users.
Through questionnaires with participants, they also confirm the clearness, conciseness, and usefulness of our generated action descriptions.

The contributions of this paper are as follows:
\begin{itemize}
    \item This is the first work to generate the caption of bug recordings to support developers in reproducing bugs.
    \item The first systematic approach \tool, to non-intrusively segment recordings into clips, infer fine-grained user actions, and create action descriptions as subtitles, with examples in online appendix\footnote{\url{https://github.com/sidongfeng/CAPdroid}}.
    \item A comprehensive evaluation including automated experiments and a user study to demonstrate the accuracy and usefulness of our approach.
\end{itemize}

\section{\tool Approach}
\label{sec:approach}
\begin{figure}  
	\centering 
	\includegraphics[width=0.95\linewidth]{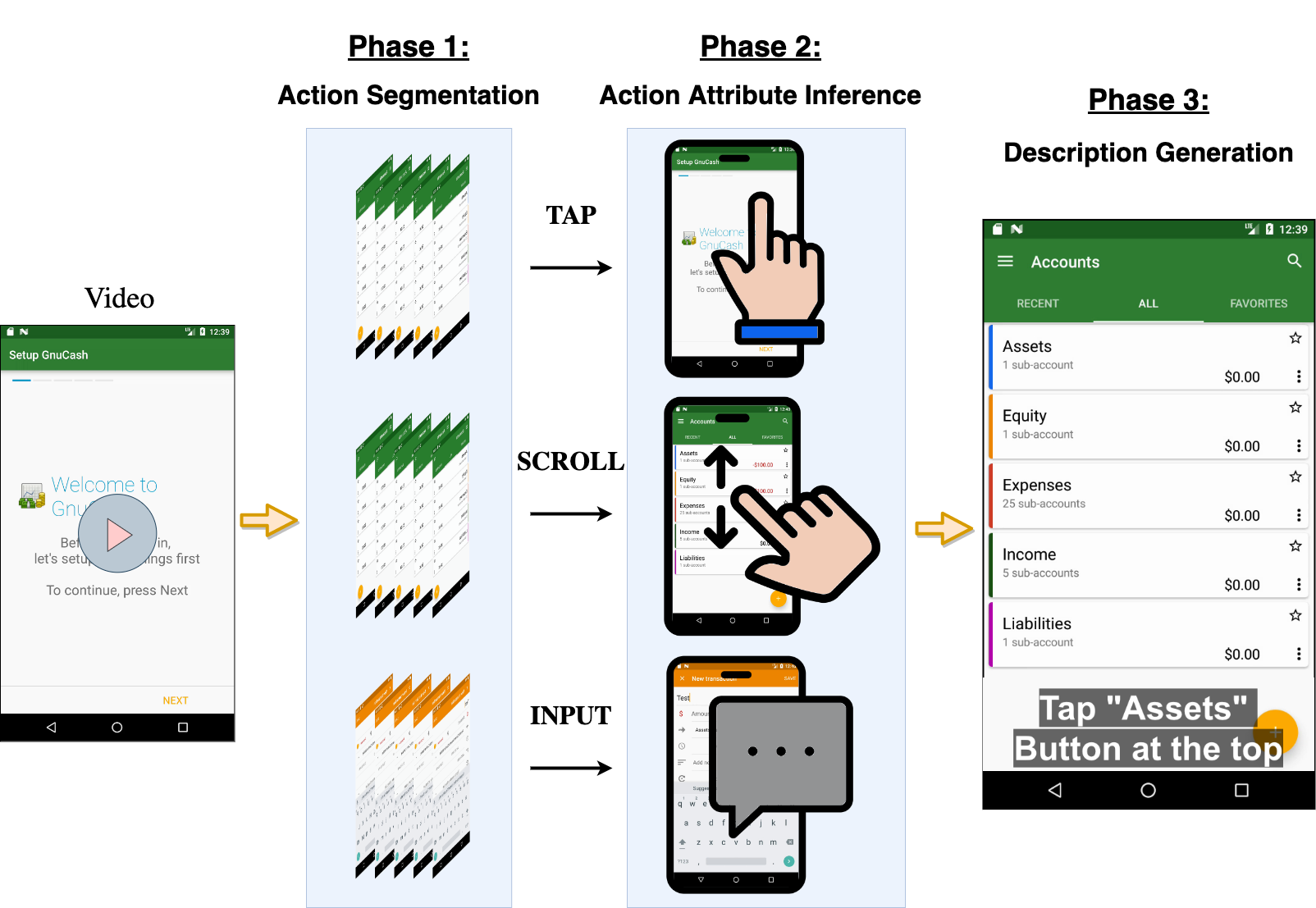}
	\caption{The overview of \tool.}
	\label{fig:overview}
\end{figure}

Given an input GUI recording, we propose an automated approach to segment the recording into a sequence of clips based on user actions and subsequently localize the action positions to generate natural language descriptions. 
The overview of our approach is shown in Fig.~\ref{fig:overview}, which is divided into three main phases: 
(i) the \textit{Action Segmentation} phase, which segments user actions from GUI recording into a sequence of clips, 
(ii) the \textit{Action Attribute Inference} phase that infers touch location, moving offset, and input text from action clips,
and (iii) the \textit{Description Generation} phase that utilizes the off-the-shelf GUI understanding models to generate high-level semantic descriptions.
Before discussing each phase in detail, we discuss some preliminary understanding of user actions in GUI recording.

\subsection{Preliminary Study}
\label{sec:background}
To understand the recordings from the end-users, we conducted a small pilot study of the GUI recordings from GitHub~\cite{web:github}.
In detail, we built a crawler to automatically crawl the bug reports from GitHub issue repositories that contain GUI recordings with suffix names like .gif, .mp4, etc.
To study more recent GUI recordings, we obtained the recordings from 2021.
Overall, we obtained 5,231 GUI recordings from 1,274 apps.
We randomly sampled 1,000 (11.5\%) GUI recordings, and we recruited two annotators online to manually check the user actions from the recordings.

Two students were recruited by the university’s internal slack channel and they were compensated with \$12 USD per hour. They have annotating experience on GUI-related (e.g., GUI element bounding box) and video-related (e.g., video classification) datasets.
To ensure accurate annotations, the process started with initial training. First, we gave them an introduction to our study and also an example set of annotated screen recordings where the labels have been annotated by the authors. Then, we asked them to pass an assessment test.
Two annotators were assigned the experimental set of screen recordings to label the user actions independently without any discussion. After the initial labeling, the annotators met and sanity corrected the subtle discrepancies. Any disagreement was handed over to the first author for the final decision.

We observed that 89\% of the recordings included a touch indicator, indicating it as a mechanism for the end-user to depict their actions on the screen.
We further classified those touch indicators into three categories, following the Card Sorting~\cite{spencer2009card} method:

\begin{itemize}
    \item \textbf{default (68\%).} As shown in Fig.~\ref{fig:default}, the touch indicator renders a small semi-transparent circle, that gives visual feedback when the user presses his finger on the device screen. This is the default touch indicator on Android.
    \item \textbf{cursor (27\%).} As shown in Fig.~\ref{fig:cursor}, users/developers may test the apps in the emulator and directly record the desktop, so that the user actions are captured by the desktop cursor.
    \item \textbf{custom (5\%).} As shown in Fig.~\ref{fig:custom}, the touch indicator is customized by third-party screen recorders, such as DU Recorder~\cite{web:durecorder}, etc.
\end{itemize}

Those findings motivated us to develop a tailored approach, exploiting touch indicators to capture end-user intent, so to generate semantic captions for GUI recording.
Considering the diversity of touch indicators in the general GUI recordings, a more advanced approach to detect and infer user actions is required.

\subsection{Phase 1: Action Segmentation}
\label{sec:phase1}
A video consists of a sequence of frames to deliver the visual detail of the story for particular scenes.
Different from the recognition of discontinuities in the visual-content flow of natural-scene videos, detecting clips in the GUI recording is to infer scenes of user actions that generally display significant changes in the GUIs.
To that end, we leverage the similarity of consecutive frames to segment user actions (i.e., \textit{TAP}, \textit{SCROLL}, \textit{INPUT}) from GUI recording.

\subsubsection{Consecutive Frame Comparison}
Inspired by signal processing~\cite{feng2022gifdroid,feng2022gifdroid2}, we leverage the image processing techniques to build a perceptual similarity score for consecutive frame comparisons based on Y-Difference (or Y-Diff).
YUV is a color space usually used in video encoding, enabling transmission errors or compression artifacts to be more efficiently masked by the human perception than using a RGB-representation~\cite{chen2009compression,sudhir2011efficient}.
Y-Diff is the difference in Y (luminance) values of two images in the YUV color space, used as a major input for the human perception of motion~\cite{livingstone2002vision}.

Consider a visual recording $\big\{ f_{0}, f_{1}, .., f_{N-1}, f_{N} \big\}$, where $f_{N}$ is the current frame and $f_{N-1}$ is the previous frame.
To calculate the Y-Diff of the current frame $f_{N}$ with the previous $f_{N-1}$, we first obtain the luminance mask $Y_{N-1}, Y_{N}$ by splitting the YUV color space converted by the RGB color space.
Then, we apply the perceptual comparison metric, SSIM (Structural Similarity Index)~\cite{wang2004image}, to produce a per-pixel similarity value related to the local difference in the average value, the variance, and the correlation of luminances.
A SSIM score is a number between 0 and 1, and a higher value indicates a strong level of similarity.

\begin{figure}  
	\centering 
	\includegraphics[width=0.95\linewidth]{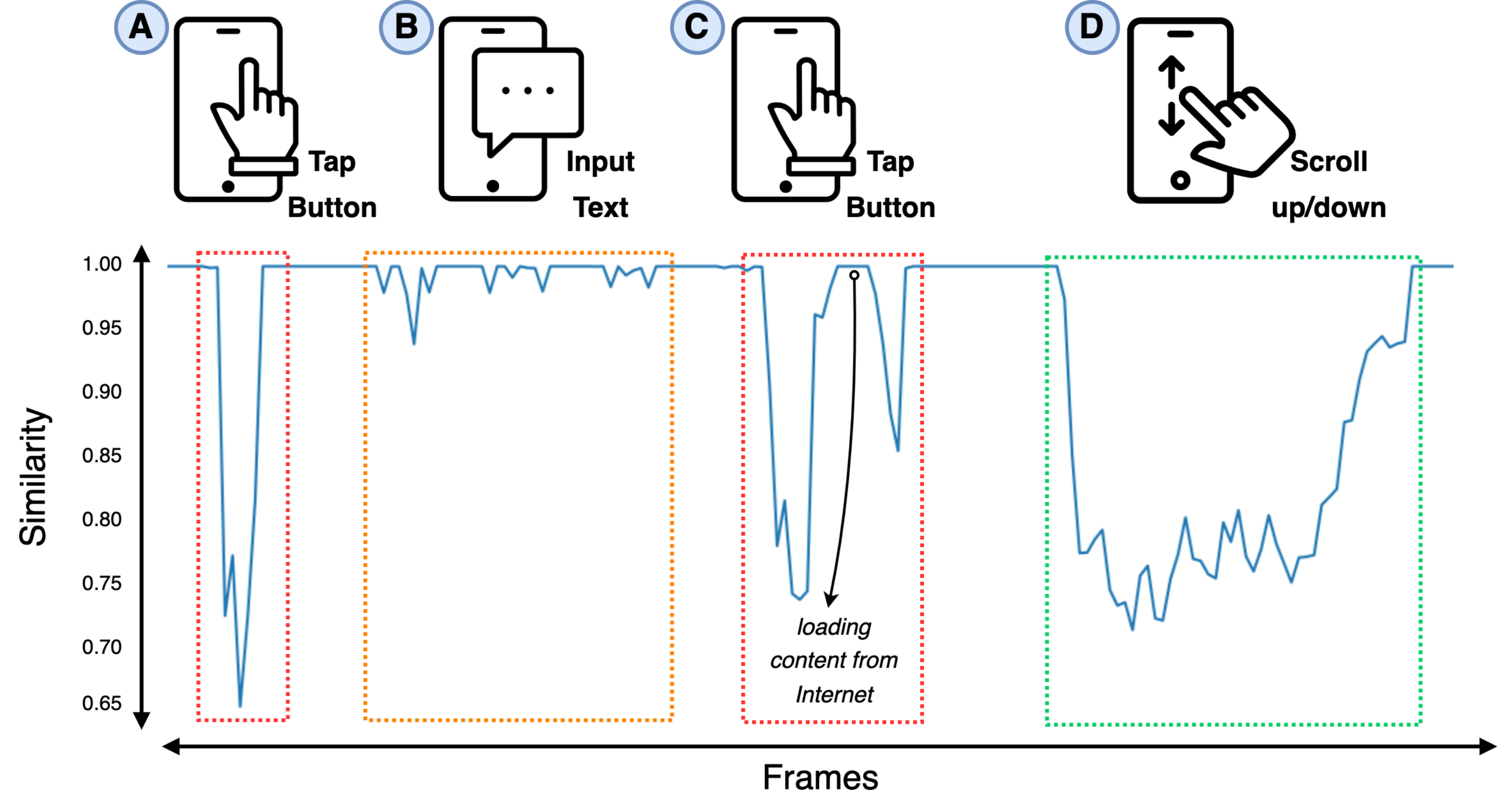}
	\caption{An illustration of consecutive frame similarity.} 
	\label{fig:timeframe}
\end{figure}

\subsubsection{Action Classification}
\label{sec:action_class}
To identify the user actions in the GUI recording, we look into the similarity scores of consecutive frames as shown in Fig.~\ref{fig:timeframe}.
The first step is to group frames belonging to the same atomic activity according to tailored pattern analysis.
This procedure is necessary because discrete activities performed on the screen will persist across several frames, and thus, need to be grouped and segmented accordingly.
Consequently, we observe three patterns of user actions, i.e., \textit{TAP}, \textit{SCROLL}, and \textit{INPUT}. 
Note that we focus on the most commonly-used actions for brevity in this paper, other actions could be extended by comparing the consecutive frame similarity.

\textit{(a) TAP}: 
As shown in Fig.~\ref{fig:timeframe}A (user taps a button), the similarity score starts to drop drastically which reveals an instantaneous transition from one screen to another.
In addition, one common case is that the similarity score becomes steady for a small period of time $t_{s}$ between two drastically droppings as shown in Fig.~\ref{fig:timeframe}C. 
The occurrence of this short steady duration $t_{s}$ is because GUI has not finished loading.
While the GUI layout of GUI rendering is fast, resource loading may take time. 
For example, rendering images from the web depends on device bandwidth, image loading efficiency, etc.

\textit{(b) SCROLL}: 
As shown in Fig.~\ref{fig:timeframe}D (user scrolls up/down the screen), the similarity score starts with a drastic drop and then continues to increase slightly over a period of time, which implicates a continuous transition from one GUI to another.

\begin{figure}
	\centering
	\subfigure[English keyboard (detect \textit{``qwert''})]{
		\includegraphics[width = 0.45\linewidth]{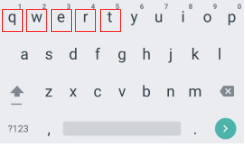}
		\label{fig:key1}}
	\hfill
	\subfigure[Numeric keypad (detect \textit{``123''})]{
		\includegraphics[width = 0.4\linewidth]{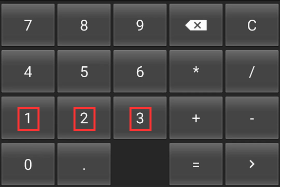}
		\label{fig:key2}}	
	\caption{Examples of keyboard detection.}
	\label{fig:key}
\end{figure}

\textit{(c) INPUT}:
As shown in Fig.~\ref{fig:timeframe}B (user inputs text), the similarity score starts to drop and rise multiple times, revealing typing characters and digits.
However, the similarity score cannot reliably detect \textit{INPUT} actions, as it may coincide with the \textit{TAP} actions.
To address this, we further supplement with Optical Character Recognition (OCR) technique~\cite{du2021pp} (a detailed description is demonstrated in Section~\ref{sec:infer_input_text}) to detect whether there is a virtual keyboard in the GUI.
Note that we focus on English apps, and it may take additional efforts to extend our approach to other languages.
In detail, we first extract the characters from the frames, and concatenate them into text-based ($ocr_{text}$) and number-based ($ocr_{num}$) string.
As the OCR may not infer the text perfectly, we discern the keyboard frame by keyboard-specific substrings.
For example, Fig.~\ref{fig:key1} is a frame of English keyboard that contains \textit{``qwert''} in $ocr_{text}$, and Fig.~\ref{fig:key2} is a frame of numeric keypad that contains \textit{``123''} in $ocr_{num}$.
Therefore, the frame of a keyboard is discriminated by
\begin{equation}
	frame = 
	\begin{cases}
		\exists \text{\{qwert, asdfg, zxcvb\}} \in lowercase(ocr_{text}) \\
		\exists \text{\{123, 456, 789\}} \in ocr_{num} \\
	\end{cases}
	\label{eq:keyboard}
\end{equation}
where $lowercase$ is to convert the uppercase characters into lowercase, in order to detect capital English keyboard.
Note that we do not adopt keyboard template matching, as keyboards vary in appearance, such as customized background, different device layouts, etc.

\begin{figure*}
	\centering
	\subfigure[TAP]{
		\includegraphics[width = 0.32\linewidth]{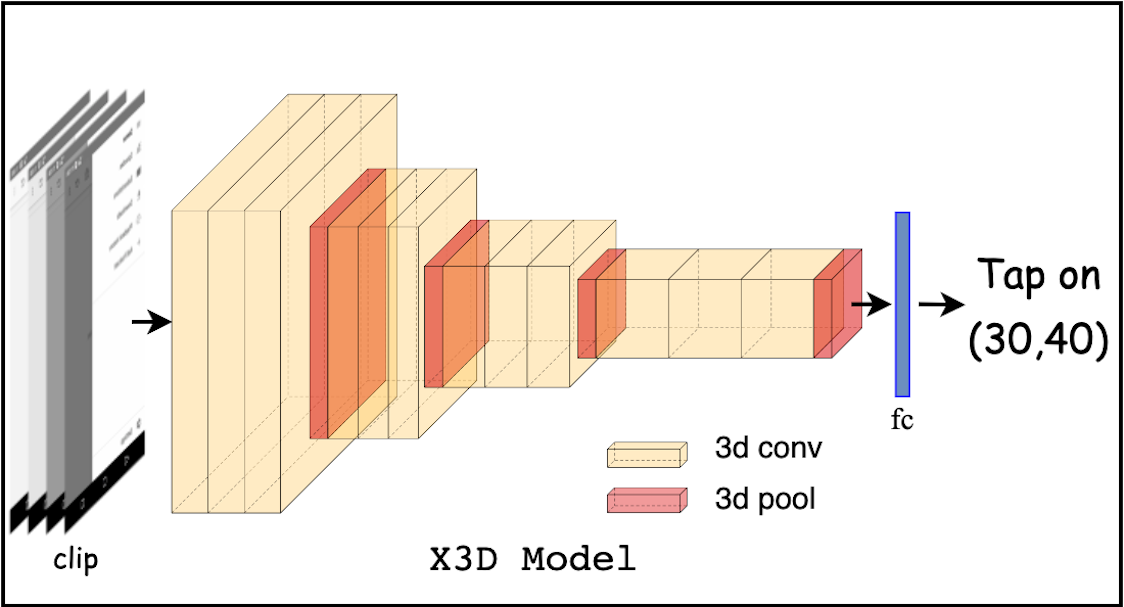}
		\label{fig:tap}}	
	\hfill
	\subfigure[SCROLL]{
		\includegraphics[width = 0.305\linewidth]{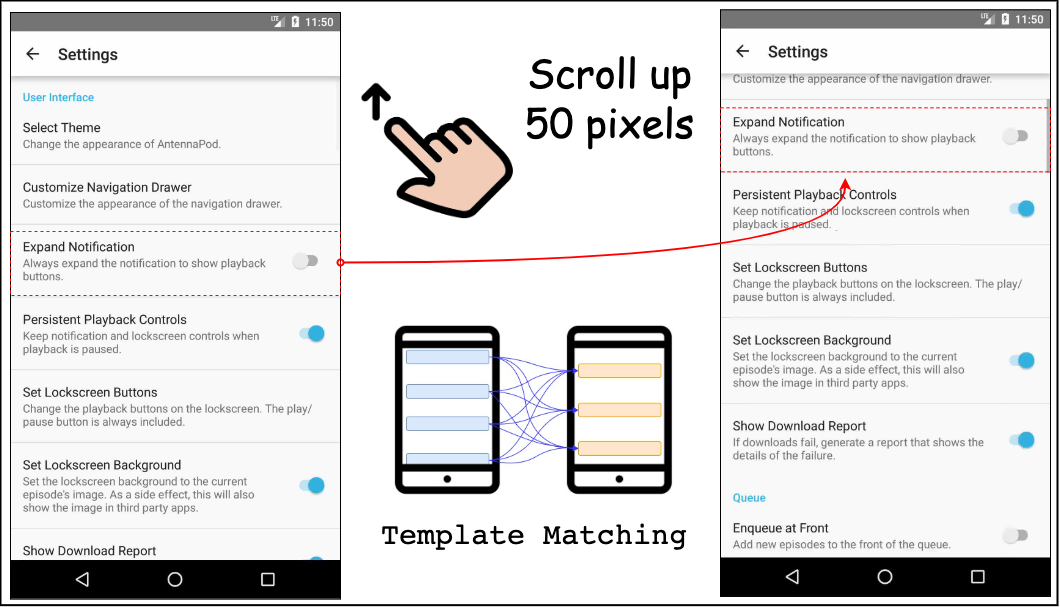}
		\label{fig:scroll}}	
	\hfill
	\subfigure[INPUT]{
		\includegraphics[width = 0.305\linewidth]{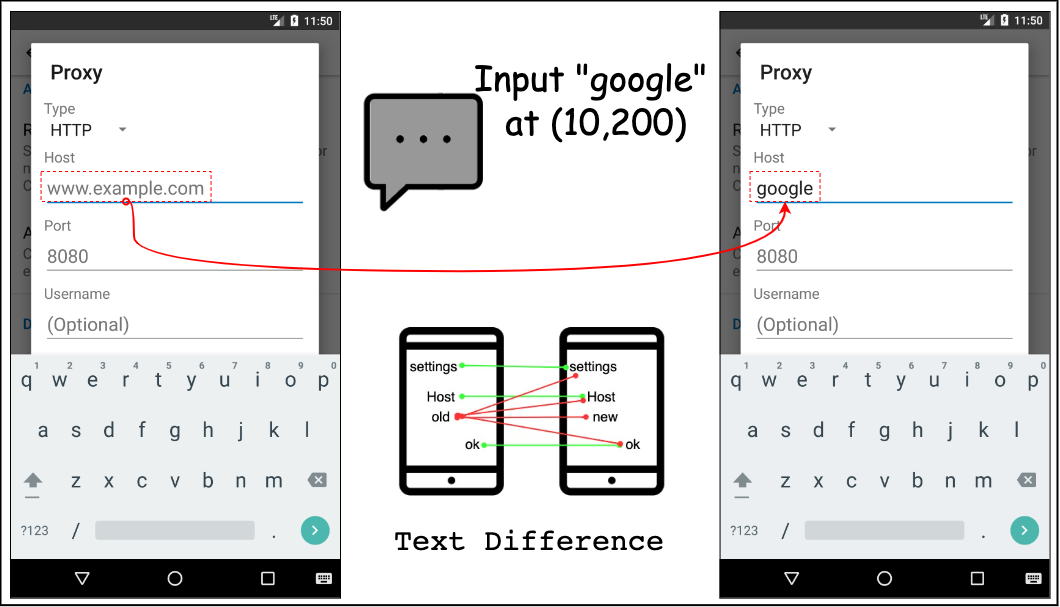}
		\label{fig:input}}	
	\caption{Approaches of Action Attribute Inference.}
	\label{fig:localization}
	\vspace{-0.3cm}
\end{figure*}

\subsection{Phase 2: Action Attribute Inference}
\label{sec:phase2}
Given a recording clip of user action segmented by the previous phase, we then infer its detailed attributes, including touch location of \textit{TAP} action, its moving offset of \textit{SCROLL} action, and its input text of \textit{INPUT} action, to reveal where the user interacts with on the screen.
The overview of our methods is shown in Fig.~\ref{fig:localization}.
The prediction of \textit{TAP} location requires a semantic understanding of the GUI transition captured in the clip, such as touch indicators (in Section~\ref{sec:background}), transition animation, GUI semantic relation, etc.
Therefore, we propose a deep-learning-based method that models the spatial and temporal features across frames to infer the \textit{TAP} location.
To infer the moving offset of \textit{SCROLL}, we adopt an off-the-shelf image-processing method to detect the continuous motion trajectory of GUIs, thus, measuring the user's scrolling direction and distance.
To infer the input text of \textit{INPUT}, we leverage the OCR technique to identify the text difference between the frames of keyboard opening (i.e., where the user starts entering text) and keyboard closing (i.e., where the user ends entering).

\subsubsection{Inferring TAP location}
\label{sec:infer_tap_location}
Convolutional Neural Networks of 2D (Conv2ds)~\cite{krizhevsky2012imagenet,lecun1998gradient} have demonstrated remarkable success in efficiently capturing the hypothesis of spatial locality in two-dimensional images.
A video that is encoded by a sequence of 2d images, aggregates another dimension: spacetime.
To predict the touch location from a GUI recording clip, we adopt a Conv3d-based model X3D~\cite{feichtenhofer2020x3d}, that simultaneously models spatial features of single-frame GUIs and temporal features of multi-frames optical flow.
The architecture of our X3D model is shown in Fig.~\ref{fig:tap}.

Given a video clip $V^{T \times H \times W \times C}$ where $T$ is the time length of the clip, $W$, $H$, and $C$ are the width, height, and channel of the frame, usually $C=3$ for RGB frame.
We first apply 3d convolution layers, consisting of a set of learnable filters to extract the  spatio-temporal features of the video.
Specifically, the convolution is to use a 3d kernel, i.e. $t \times d \times d$ where $t$ and $d$ denote the temporal and spatial kernel size, to slide around the video and calculate kernel-wise features by matrix dot multiply.
After the convolutional layers, the video $V$ will be abstracted as a 3d feature map, preserving features along both the spatial and the temporal information.
We then apply a 3d pooling layer to eliminate unimportant features and enhance spatial variance of rotation and distortion.
After blocks of convolutional and pooling layers, we flatten the feature map and apply a fully connected layer to infer the logits of \textit{TAP} location.

For the detailed implementation, we adopt the convolutional layers from ResNet-50~\cite{he2016deep} and borrow the idea of residual connection to improve the model performance and stability between layers.
We use MaxPooling~\cite{schmidhuber2015deep} as the pooling layer, where the highest value from the kernel is taken, for noise suppressant during abstraction. 
The output of the fully connected layer is 2 neurons, representing $(x,y)$ coordinates.
To accelerate the training process~\cite{ren2015faster}, we standardize the coordinate relative to the width and height of the frame.
Although the frames are densely recorded (i.e. 30fps), the GUI renders slowly.
To extract discriminative features from the recording, we uniformly sample 16 frames at 5 frame intervals ($T=16$) as suggested in~\cite{feichtenhofer2020x3d}.
Note that if the length of the recording clip is smaller than the sample rate $16 \times 5$, we will sample the frames based on nearest neighbor interpolation.
To make our training more stable, we adopt Adam as the optimizer~\cite{kingma2014adam} and MSELoss as the loss function~\cite{murphy2012machine}.
Moreover, to optimize the model, we apply an adaptive learning scheduler, with an initial rate of 0.01 and decay to half after 10 iterations.
The hyper-parameter settings are determined empirically by a small-scale experiment.


\subsubsection{Inferring SCROLL offset}
To infer the scrolling direction (i.e., upward, downward) and distance (i.e., amount of movement) from the GUI recording clip, we measure the motion trajectory of GUI elements.
Since the elements may scroll off-screen~\cite{irani2006improving}, we adopt the K-folds template matching method as shown in Fig.~\ref{fig:scroll}.

Given a GUI recording clip $\big\{ f_{0}, f_{1}, .., f_{N-1}, f_{N} \big\}$ , where $f_{N}$ is the current frame and $f_{N-1}$ is the previous frame.
We first divide the previous GUI $f_{N-1}$ into K pieces vertically.
We set K to 10 by a small pilot study to mitigate the off-screen issue and preserve sufficient features for template matching.
And then, we match the template of each fold in the current frame $f_{N}$ to compute the scrolling offset between consecutive frames. 
At the end, we derive the scrolling distance by summing the offsets ($ \sum_{n=0}^{N} \text{\textit{offset}}_n^{n-1} $), and infer the scrolling direction by the sign of the distance, e.g., positive for downward, otherwise upward.

\subsubsection{Inferring INPUT text}
\label{sec:infer_input_text}
Detecting input text based on user actions on the keyboard can be error-prone, as the user may edit text from the middle of the text, switch to capital, delete text, etc.
Therefore, we leverage a practical OCR technique PP-OCRv2~\cite{du2021pp} to detect the text difference between the first frame (opening keyboard) and the last frame (closing keyboard) from the \textit{INPUT} recording clip, as shown in Fig.~\ref{fig:input}.
Given a GUI frame, PP-OCRv2 detects the text areas in the GUI by using an image segmentation network and then applies a sequence and classification model to recognize the text.
As the GUI text is similar to scene text~\cite{chen2020object}, we directly use the pre-trained PP-OCRv2 without any fine-tuning on GUI text, that the overall performance reaches 84.3\% state-of-the-art accuracy.


After deriving the text from the frames of keyboard opening and keyboard closing, we first remove all the text on the keyboard to keep the text concise.
Then, we detect the text difference between the frames using SequenceMatcher~\cite{web:difflib}.
Albeit good performance of PP-OCRv2, it may still make wrong text recognition, e.g., missing space.
To address this, SequenceMatcher measures text similarity by computing the longest contiguous matching subsequence (LCS).
Finally, we extract the text that appears only in the frame where the keyboard is closed, as input text.

\renewcommand{\arraystretch}{1.2}
\begin{table*}
    \small
	\footnotesize
	\caption{Description template, where ``\textit{obj}'' and ``\textit{nbr}'' denote the GUI element and its neighbor, $\alpha$ and $\beta$ denote high- and low-confidence element.}
	\label{tab:template}
	\begin{tabularx}{\textwidth}{l|c|l|l|X} 
		\hline
		\bf{Action} & \bf{Id} & \bf{Condition} & \bf{Template} & \bf{Example} \\
		\hline
		\hline
		\multirow{5}{*}{\textit{TAP}} & 1 & (\textit{obj\textsubscript{text/caption}} $\neq$ \textit{NULL}) $\land$ (\textit{obj\textsubscript{confid}} $> \alpha$) & Tap [\textit{obj\textsubscript{text}}] [\textit{obj\textsubscript{class}}] & Tap ``OK'' button \\
		\cline{2-5}
         & 2 & (\textit{obj\textsubscript{text/caption}} $\neq$ \textit{NULL}) $\land$ ($\beta <$ \textit{obj\textsubscript{confid}} $< \alpha$) & Tap [\textit{obj\textsubscript{text}}] [\textit{obj\textsubscript{class}}] at [\textit{obj\textsubscript{position}}] & Tap ``menu'' icon at top left corner \\
        \cline{2-5}
         & 3 & (\textit{obj\textsubscript{text/caption}} == \textit{NULL}) $\lor$ (\textit{obj\textsubscript{confid}} $< \beta$) & Tap the [\textit{obj\textsubscript{class}}] [\textit{nbr\textsubscript{relation}}] [\textit{nbr\textsubscript{text}}] & Tap the checkbox next to ``Dark Mode'' \\
		\hline
		\multirow{2}{*}{\textit{SCROLL}} & 4 & \textit{obj\textsubscript{text}} $\neq$ \textit{NULL} & Scroll [\textit{direction}] [\textit{offset}] of the screen to [\textit{obj\textsubscript{text}}] & Scroll down half of the screen to ``Advanced Setting'' \\
		\cline{2-5}
		 & 5 & \textit{obj\textsubscript{text}} == \textit{NULL} & Scroll [\textit{direction}] [\textit{offset}] of the screen & Scroll up a quarter of the screen \\
		\hline
		\multirow{3}{*}{\textit{INPUT}} & 6 & (\textit{obj\textsubscript{text/caption}} $\neq$ \textit{NULL}) $\land$ (\textit{obj\textsubscript{confid}} $> \alpha$) & Input [\textit{text}] in the [\textit{obj\textsubscript{text}}] edittext & Input ``100'' in the ``Amount'' edittext \\
		\cline{2-5}
		 & 7 & (\textit{obj\textsubscript{text}} == \textit{NULL}) $\lor$ (\textit{obj\textsubscript{confid}} $< \alpha$) & Input [\textit{text}] in the edittext [\textit{nbr\textsubscript{relation}}] [\textit{nbr\textsubscript{text}}] & Input ``John'' in the edittext below ``Name'' \\
		\hline
	\end{tabularx}
\end{table*}

\begin{figure}  
	\centering 
	\includegraphics[width=0.65\linewidth]{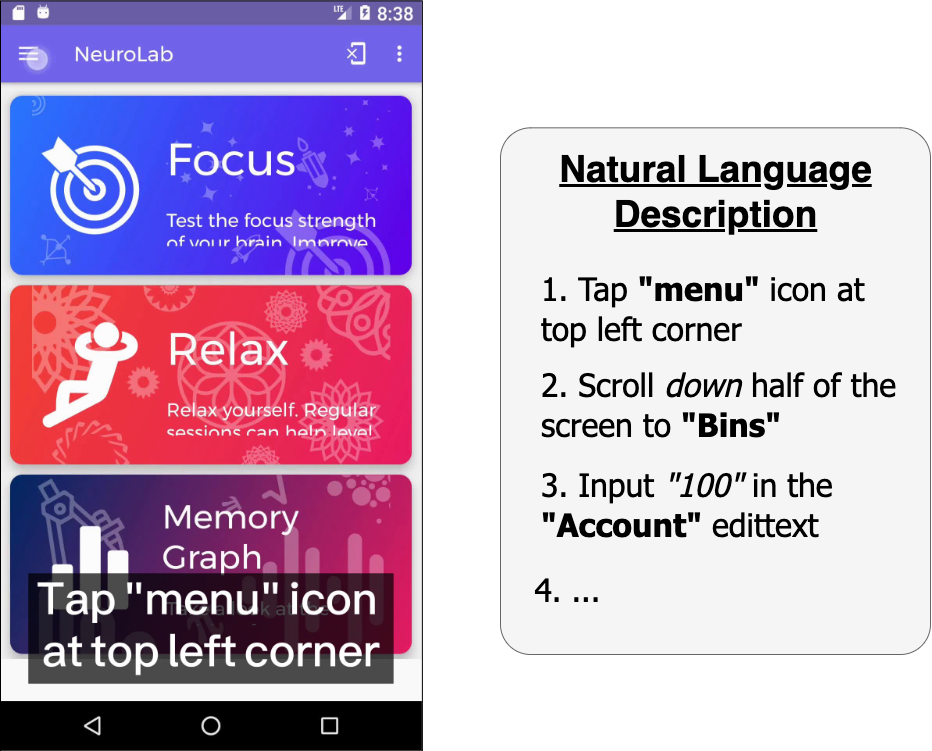}
	\caption{Subtitle and textual steps in the GUI recording.} 
	\label{fig:subtitle}
\end{figure}

\subsection{Phase 3: Description Generation}
\label{sec:phase3}
Once the attributes of the action are derived from the previous phases, we proceed by generating in-depth and easy-to-understand natural language descriptions.
To accomplish this, we first leverage mature GUI understanding models to obtain GUI information non-intrusively.
Then, we propose a novel algorithm to phrase actions into descriptions and embed them as subtitles in the recording as shown in Fig.~\ref{fig:subtitle}.

\subsubsection{GUI understanding} 
\label{sec:gui_understanding}

\begin{figure}  
	\centering 
	\includegraphics[width=0.75\linewidth]{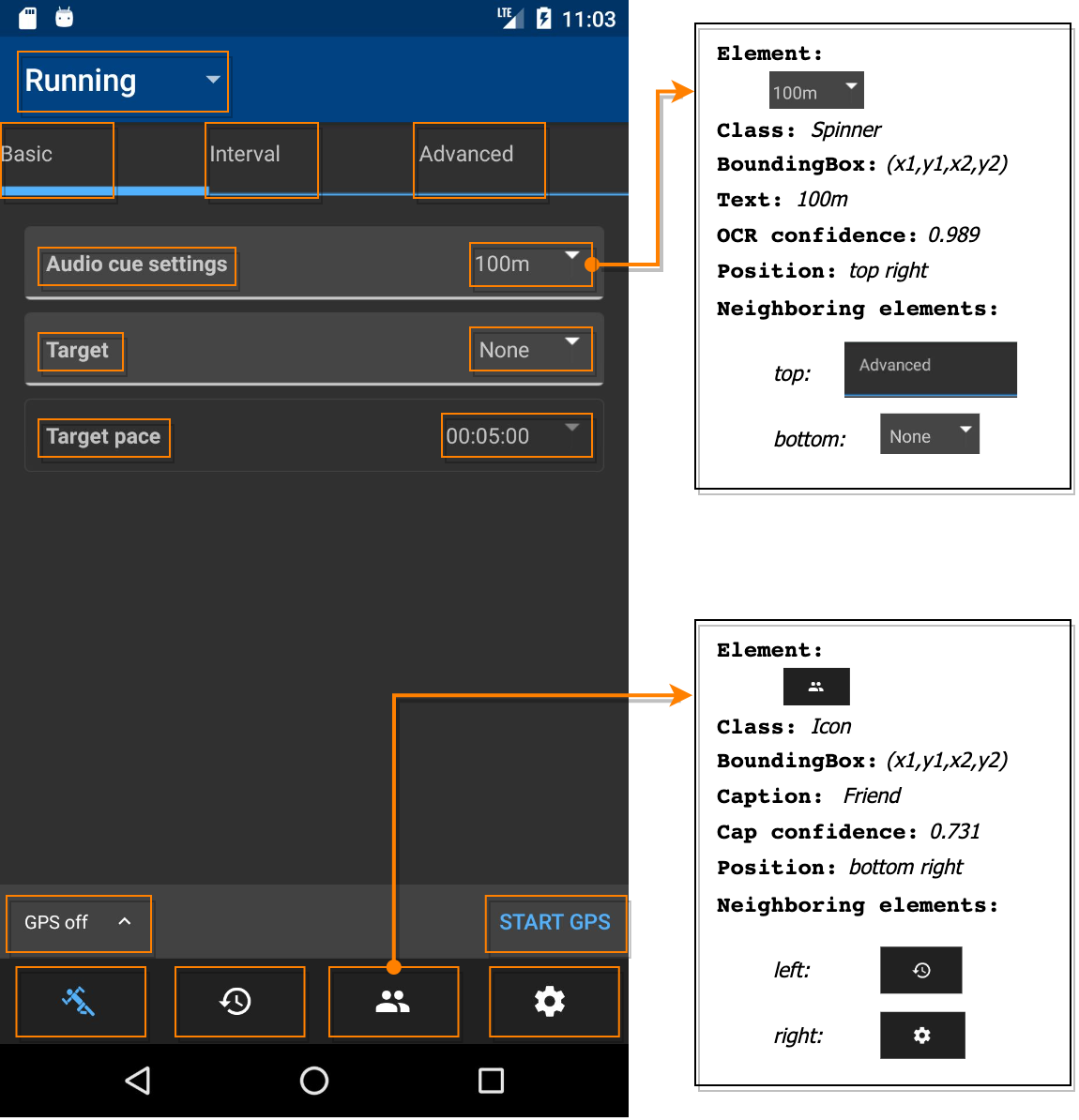}
	\caption{Example of GUI understanding.} 
	\label{fig:understanding}
\end{figure}

To understand the GUI, we adopt non-intrusive approaches to obtain GUI information, to avoid the complexity of app instrumentation or handling of the diverse software stack, especially for closed-source systems where no underlying instrumentation support is accessible~\cite{qian2020roscript}.
An example of GUI understanding is shown in Fig.~\ref{fig:understanding}.

Specifically, we first implement the state-of-the-art object detection model Faster-RCNN with ResNet-101~\cite{he2016deep} and Feature Pyramid Networks~\cite{lin2017feature} to detect 11 GUI element classes on the screen:
button, checkbox, icon, imageview, textview, radio button, spinner, switch, toggle button, edittext, and chronometer.
We train the model on the Rico dataset~\cite{deka2017rico} contains 66k GUIs from 9.7k apps. Following the previous work~\cite{zhang2021screen}, we split the GUIs in the training:validation:testing dataset by apps in the ratio of 8:1:1.
As a result, the model achieves an overall Mean Average Precision (MAP) of 51.45\% on the test set.
For each GUI element, we adopt the OCR technique (the detailed implementation is elaborated in Section~\ref{sec:infer_input_text}) to detect the text (if any). 
For the icon, annotation based on common human understanding can enhance the GUI understanding.
For example, in Fig.~\ref{fig:understanding}, the icon of a group of people informs the semantic of ``\textit{Friend}''.
To achieve this, we adopt a transformer-based model from the existing work~\cite{chen2020unblind} to caption the icon image.
We follow the implementation in their original paper to train the model and achieve 60.7\% accuracy on the test set.

Besides from understanding the information of GUI elements, we also attempt to obtain their global information relative to the GUI, including absolute positioning and element relationship.
Absolute positioning describes the element as a spatial position in the GUI, which is particularly useful to represent an element in an image~\cite{parmar2018image}.
To accomplish this, we uniformly segment the GUI into $3 \times 3$ grids, delineating horizontal position (i.e, \textit{left, right}), vertical position (i.e, \textit{top, bottom}), and \textit{center} position.
For example, in Fig.~\ref{fig:understanding}, the ``100m'' spinner is at \textit{the top right} corner.
GUI element relationship aims to transform the ``flat'' structure of GUI elements into connected relationships.
A natural way of representing the relationship is using a graph structure, where elements are linked to the nearest elements. 
To accomplish this, we first compute the horizontal and vertical distance between GUI elements by euclidean pixel measurement.
And then, we construct the graph of the GUI elements by finding the nearest elements (neighbors) in four directions, including \textit{left, right, top, and bottom}.
Note that we set up a threshold to prevent the neighbors from being too far apart.
Ultimately, it will generate a graph representing the relationships between the elements in the GUI.
For example, in Fig.~\ref{fig:understanding}, the ``100'' spinner has two neighbors: the ``Advanced'' element at the \textit{top}, and the ``None'' element at the \textit{bottom}.
Note that the ``Audio cue settings'' element is omitted due to large spacing, which is consistent with human viewing.

\subsubsection{Subtitle Creation}
The main instruction of interest is to create a clear and concise subtitle description based on $\big\{ action, object \big\}$.
The global GUI information is further used to complement the description by $\big\{ position, relationship \big\}$.
Based on the $action$ obtained in Section~\ref{sec:phase1}, the attribute of $object$ inferred in Section~\ref{sec:phase2}, and the corresponding GUI element information retrieved in Section~\ref{sec:gui_understanding}, we propose description templates for \textit{TAP}, \textit{SCROLL}, \textit{INPUT}, respectively.
A summary of description templates can be seen in Table~\ref{tab:template}.

For \textit{TAP} action, the goal of the description should be clear and concise, e.g., tap ``OK'' button.
However, we find that this simple description may not articulate all \textit{TAP} actions due to two reasons.
First, the text and caption of \textit{object} are prone to errors or undetected, as the OCR-obtained text and the caption-obtained annotation are not 100\% accurate.
Second, there may be multiple \textit{objects} with the same text on the GUI.
To resolve this, we set up an \textit{object} confidence value \textit{obj\textsubscript{confid}} as:
\begin{equation}
\textit{obj\textsubscript{confid}} = 
	\begin{cases}
		\textit{OCR\textsubscript{confid}} & \text{if \textit{obj\textsubscript{text}} is unique in GUI}\\
		0 & \text{otherwise}
	\end{cases}
	\label{eq:orb_descriptor}
\end{equation}
where \textit{OCR\textsubscript{confid}} denotes the confidence predicted by OCR.
Note that the confidence value of icon \textit{object} is calculated likewise by captioning.
The smaller the confidence value, the less intuitive the \textit{object} is.
Therefore, only the \textit{object} with the highest confidence value (\textit{obj\textsubscript{confid}} $> \alpha$) will apply the simplest and most straightforward description (Template 1), otherwise, we add the context of absolute position to help locate the \textit{object} (Template 2).
For the \textit{object} whose text is not detected or recognized with low confidence, we leverage the context of its \textit{neighbor} to help locate the target \textit{object} (Template 3), e.g., tap the checkbox next to ``Dark Mode''.

It is easy to describe a \textit{SCROLL} action by its scrolling direction and offset (Template 5), e.g., scroll up a quarter of the screen.
However, such an offset description is not precise and intuitive.
To address this, if a new element with text appears by scrolling, we add this context to help describe where to scroll to (Template 4), e.g., scroll down half of the screen to ``Advanced Setting''.

The description of \textit{INPUT} is similar to \textit{TAP}.
For the high-confidence \textit{object} with text (Template 6), it generates: Input [\textit{text}] in the [\textit{obj\textsubscript{text}}] edittext.
Different from the \textit{TAP} descriptions, we do not apply the context of absolute position to help locate the low-confidence \textit{object}.
This is because the \textit{objects} are gathering at the top when the keyboard pops up, so the absolute positioning may not help.
Instead, we use the relative position of \textit{neighbor} to describe the input \textit{object} of which text is not detected or recognized with low confidence (Template 7), e.g., Input ``John'' in the edittext below ``Name''.

After generating the natural language description for each action clip, we embed the description into the recording as subtitles as shown in Fig.~\ref{fig:subtitle}.
In detail, we create the subtitles by using the Wand image annotation library~\cite{web:wand} and synchronize the subtitle display at the beginning of each action clip.

\section{Automated Evaluation}
\label{sec:evaluation}
In this section, we described the procedure we used to evaluate \tool in terms of its performance automatically.
Since our approach consists of two main automated steps to obtain the actions from the recordings, we evaluate these phases accordingly, including Action Segmentation (Section~\ref{sec:phase1}), and Action Attribute Inference (Section~\ref{sec:phase2}).
Consequently, we formulated the following two research questions:
\begin{itemize}
    \item \textbf{RQ1}: How accurate is our approach in segmenting action clips from GUI recordings?
	\item \textbf{RQ2}: How accurate is our approach in inferring action attributes from clips?
\end{itemize}

To perform the evaluation automatically, we leveraged the existing automated app exploration tool Droidbot~\cite{li2017droidbot} to collect GUI recordings with ground-truth actions.
In detail, we first collected 439 top-rated Android apps from Google Play covering 14 app categories (e.g., news, tools, finance, etc.). 
Each app was run for 10 minutes by Droidbot to automatically explore app functionalities by simulating user actions on the GUI.
The simulated actions, including operation time, types, locations, etc, were dumped as metadata, representing the ground truth.
Meanwhile, we captured a screen recording to record the actions for each app at 30 fps.
As discussed in Section~\ref{sec:background}, users may use different indicators to depict their touches.
To make our recordings as similar to real-world recordings as possible, we adopted different touch indicators to record actions, including 181 default, 152 cursor, and 106 custom.
In total, we obtained 439 10-min screen recordings as the experimental dataset for the evaluation.

\subsection{RQ1: Accuracy of Action Segmentation}
\label{sec:evaluation_1}
\textbf{Experimental Setup.}
To answer RQ1, we evaluated the ability of our \tool to precisely segment the recordings into action clips and accurately classify the actions.
To accomplish this, we utilized the metadata of action operation time as the ground-truth.
During preliminary observation with many recordings, we found that, due to the delay between commands and operations on the device, it may have small time-frame differences between the ground-truth and the recorded actions.
To avoid these small differences, we broadened the ground-truth of the actions by 5 frames.
In total, we obtained 12k \textit{TAP}, 4k \textit{SCROLL}, and 1k \textit{INPUT} clips from 439 screen recordings.


\textbf{Metrics.}
We employed two widely-used evaluation metrics, e.g., video segmentation F1-score, and accuracy.
To evaluate the precision of segmenting the action clips from recordings, we adopted video segmentation F1-score~\cite{truong2021automatic}, which is a standard video segmentation metric to measure the difference between two sequences of clips that properly accounts for the relative amount of overlap between corresponding clips.
Consider the clips segmented by our method ($c_{our}$) and ground truth ($c_{gt}$), vs-score is computed as $ \frac{2|c_{our} \cap c_{gt}|}{|c_{our}| + |c_{gt}|} $, where $|c|$ denotes the duration of the clip.
The higher the score value, the more precise the method can segment the video.
We further adopted accuracy to evaluate the performance of our approach to discriminate action types from clips.
The higher the accuracy score, the better the approach can classify the actions.

\textbf{Baselines.}
To demonstrate the advantage of using SSIM as the image similarity metric to segment actions from GUI recordings, we compared it with 5 image-processing baselines, including pixel level (e.g, absolute differences ABS~\cite{watman2004fast}, color histogram HIST~\cite{wang2010robust}), structural level (e.g., SIFT~\cite{lowe2004distinctive}, SURF~\cite{bay2006surf}), and motion-estimation level (e.g., edge detection EDGE~\cite{zhan2007improved}).
Due to the page limit, we omitted the details of these well-known methods.

\renewcommand{\arraystretch}{1.05}
\begin{table}
    \footnotesize
	\centering
	\caption{Performance comparison of action segmentation. ``VS'' denotes the video segmentation F1-score, and ``Acc'' denotes the accuracy of action classification. }
	\label{tab:segmentation_performance}
	\begin{tabular}{l|c|c|c|c|c|c||c|c} 
	    \hline
	    \multirow{2}{*}{\bf{Method}} & \multicolumn{2}{c|}{\bf{TAP}} & \multicolumn{2}{c|}{\bf{SCROLL}} & \multicolumn{2}{c||}{\bf{INPUT}} & \multicolumn{2}{c}{\bf{Overall}} \\
	    \cline{2-9}
	     & VS & Acc & VS & Acc & VS & Acc & VS & Acc\\
	     \hline
	    ABS & 0.56 & 0.69 & 0.59 & 0.69 & 0.67 & 0.73 & 0.61 & 0.71 \\
	    HIST & 0.71 & 0.80  & 0.62 & 0.71 & 0.75 & 0.84 & 0.70 & 0.79 \\
	    SIFT & 0.61 & 0.71 & 0.60 & 0.73 & 0.63 & 0.79 & 0.62 & 0.75 \\
	    SURF & 0.55 & 0.71 & 0.59 & 0.72 & 0.60 & 0.77 & 0.58 & 0.74 \\
	    EDGE & 0.61 & 0.75 & 0.55 & 0.70 & 0.66 & 0.78 & 0.61 & 0.75 \\
	    \bf{Ours} & \bf{0.81} & \bf{0.89} & \bf{0.83} & \bf{0.92} & \bf{0.90} & \bf{0.97} & \bf{0.84} & \bf{0.93} \\
		\hline
	\end{tabular}
	\vspace{-0.1cm}
\end{table}

\textbf{Results.}
Table~\ref{tab:segmentation_performance} shows the overall performance of all baselines.
The performance of our method is much better than that of other baselines, i.e., 20\%, 17\% boost in video segmentation F1-score and accuracy compared with the best baseline (HIST).
Although HIST achieves the best performance in the baselines, it does not perform well as it is sensitive to the pixel value.
This is because the recordings can often have image noise due to fluctuations of color or luminance.
The image similarity metrics based on structural level (i.e., SIFT, SURF) are not sensitive to image pixel, however, they are not robust to compare GUIs.
This is because, unlike images of natural scenes, features in the GUIs may not distinct.
For example, a GUI contains multiple identical checkboxes, and the duplicate features of checkboxes can significantly affect similarity computation.
Besides, motion-estimation baseline (EDGE) cannot work well in segmenting actions from GUI recordings, as GUI recordings are artificial artifacts with different rendering processes.
In contrast, our method using SSIM achieves better performance as it takes similarity measurements in many aspects from spatial and pixel, which allows for a more robust comparison.

\renewcommand{\arraystretch}{1}
\begin{table*}
    \small
	\centering
	\caption{Performance comparison of action attribute inference.}
	\label{tab:localization_performance}
	\begin{tabular}{l|c|c|c|c|c|c|c|c|c|c} 
	    \hline
	    \multirow{2}{*}{\bf{Methods}} & \multicolumn{3}{c|}{\bf{TAP}} & \multicolumn{3}{c|}{\bf{SCROLL}} & \multicolumn{3}{c|}{\bf{INPUT}} & \multirow{2}{*}{\bf{Overall}} \\
	    \cline{2-10}
	     & default & cursor & custom & default & cursor & custom & default & cursor & custom & \\
	     \hline
	    V2S~\cite{bernal2020translating} & 84.19\% & 69.66\% & 36.10\% & 85.19\% & 63.31\% & 29.00\% & - & - & - & 61.24\% \\
	    GIFdroid~\cite{feng2022gifdroid} & 85.78\% & 88.01\% & 87.16\% & 72.84\% & 71.01\% & 69.77\% & 35.13\% & 32.11\% & 28.39\% & 63.35\% \\
	    \bf{\tool} & \bf{91.06\%} & \bf{90.28\%} & \bf{92.67\%} & \bf{94.87\%} & \bf{94.63\%} & \bf{95.12\%} & \bf{87.86\%} & \bf{88.62\%} & \bf{88.11\%} & \bf{91.46\%} \\
		\hline
	\end{tabular}
\end{table*}

Our method also makes mistakes in action segmentation due to two reasons.
First, we wrongly segment one action clip into multiple ones due to the unexpected slow resource loading, e.g., one clip for the GUI transition of a user action, and the other clip for the GUI's resource loading.
Second, some GUIs may contain animated app elements such as advertisements or movie playing, which will change dynamically, resulting in mistake action segmentation and classification.

\subsection{RQ2: Accuracy of Action Attribute Inference}
\label{sec:evaluation_2}

\textbf{Experimental Setup.}
To answer RQ2, we evaluated the ability of our \tool to accurately infer the action attributes from the segmented clips.
To accomplish this, we leveraged the metadata of action attributes as the ground-truth.
Since our approach employs a deep-learning-based model (Section~\ref{sec:infer_tap_location}) to infer \textit{TAP} location, we trained and tested our model based on the metadata of \textit{TAP} actions.
Note that a simple random split cannot evaluate the model's generalizability, as tapping on the screens in the same app may have very similar visual appearances.
To avoid this data leakage problem~\cite{kaufman2012leakage}, we split the \textit{TAP} actions in the dataset by apps, with the 8:1:1 app split for the training, validation, and testing sets, respectively.
We also ensure a similar number of three types of touch indicators (i.e. default, cursor, custom) in the split dataset.
The resulting split has 9k actions in the training dataset, 1.5k in the validation dataset, and 1.5k in the testing dataset.
The model was trained in an NVIDIA GeForce RTX 2080Ti GPU (16G memory) with 30 epochs.
In total, we obtained 1.5k \textit{TAP} locations, 4k \textit{SCROLL} offsets, and 1k \textit{INPUT} text as the attributes of testing data. 

\textbf{Metrics.}
We employed accuracy as the evaluation metric to measure the performance of our approach in inferring \textit{TAP}, \textit{SCROLL}, and \textit{INPUT} action attributes, respectively.
As one element occupies a certain area, tapping any specific point within that area can successfully trigger the action.
So, we measured whether our predictions are within the ground-truth element.
For \textit{SCROLL} actions, we measured whether our inferred scroll offset is the same as the ground-truth.
For \textit{INPUT} actions, we measured whether our approach can infer the correct input text.
The higher the accuracy score, the better the approach to infer action attributes.

\textbf{Baselines.}
We set up 2 state-of-the-art methods as our baselines to compare with our \tool.
\textit{V2S}~\cite{bernal2020translating} proposed the first GUI video analysis technique, that utilizes deep-learning models to detect the touch indicator for each frame in a video and then classify them to user actions.
As V2S only detects the default touch indicator, we followed their procedure to train corresponding deep-learning models to detect cursor and custom indicators.
\textit{GIFdroid}~\cite{feng2022gifdroid} developed a novel lightweight tool to detect the user actions by first extracting the keyframes from the GUI recording and then mapping it to the GUI transition graph (UTG) to extract the execution actions.
We also followed the details in their paper to obtain the UTG graph.

\begin{figure}  
	\centering 
	\includegraphics[width=0.85\linewidth]{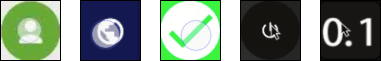}
	\caption{Examples of bad cases in action localization.} 
	\label{fig:false}
\end{figure}

\textbf{Results.}
Table~\ref{tab:localization_performance} shows the overall performance of all methods.
Our method outperforms in all actions, e.g., on average 91.33\%, 94.87\%, 88.19\% for \textit{TAP}, \textit{SCROLL}, and \textit{INPUT}, respectively.
Our method is on average 30.2\% more accurate compared with V2S in action attribute inference, due to three main reasons.
First, our method models the features from both the spatial (i.e., touch indicator) and temporal (i.e., GUI animation) across the frames to enhance the performance of the model in inferring \textit{TAP} actions, i.e., on average 91.33\% vs 63.32\% for \tool and V2S respectively.
Second, our method achieves better performance in inferring action attributes even for the recordings with different touch indicators.
This is because, \tool proposes a novel touch indicator-independent method by leveraging the similarity of consecutive frames to identify actions, while V2S leverages the opacity of the indicator, e.g., a fully solid touch indicator represents the user first touches the screen, and it fades to less opaque when a finger is lifted off the screen.
The opacity of the indicator works well for the default touch indicator (on average 84.69\%), but not for the others (on average 66.48\%, 32.55\% for cursor and custom).
Third, \tool can accurately (on average 88.19\%) infer the input text from the clips, while V2S cannot detect semantic actions.

\tool is on average 28\% (91.46\% vs 63.35\%) more accurate even compared with the best baseline (GIFdroid).
This is because, the content in GUIs of some apps (e.g., financial, social, music apps) are dynamic, causing the keyframes wrongly map to the states in the UTG.
This issue further exacerbates input text inference, as the input text from the recording is specific but the input text in UTG is randomly generated. 

Albeit the good performance of our approach, we still make wrong inferences about some actions. 
We manually check those wrong cases and find two common causes.
First, as shown in Fig.~\ref{fig:false}, the overlap of similar colors between the touch indicators and icons leads to less distinct features of the indicators, causing false-positive action localization.
Second, although the good performance of our OCR method, it still makes wrong text recognition, especially missing spaces.
We believe the emergence of advanced OCR methods can further improve the accuracy of our approach.

\section{Usefulness Evaluation}
\label{sec:user_study}
In this section, we conducted a user study to evaluate the usefulness of our generated descriptions (reproduction steps) for replaying bug recordings in real-world development environments. 

\textbf{Procedure:}
We recruited another 8 participants including 6 graduate students (4 Master, 2 Ph.D) and 2 software developers to participate in the experiment. 
All students have at least one-year experience in developing Android apps and have worked on at least one Android apps project as interns in the company.
Two software developers are more professional and have two-year working experience in a large company in Android development.
Given that they all have experience in Android app development and bug replay, they are recognized as substitutes for developers in software engineering research experiments~\cite{salman2015students}.

To mitigate the threat of user distraction, we conducted the experiment in a quiet room individually without mutual discussion.
We first gave them an introduction to our study and also a real example to try.
Each participant was then asked to reproduce the same set of 10 randomly selected bug recordings from real-world issue reports in GitHub, on average, 3.6 TAP, 1.2 SCROLL, and 1.0 INPUT per recording.
The experimental bug recordings can be seen in our online appendix\footnote{\url{https://github.com/sidongfeng/CAPdroid}}.
The study involved two groups of four participants: the control group $P_1$, $P_2$, $P_3$, $P_4$ who gets help with the reproduction steps written by reporters from GitHub, and the experimental group $P_5$, $P_6$, $P_7$, $P_8$ who gets help with the natural language description generated by our tool.
Each pair of participants $\langle P_x$, $P_{x+4}\rangle$ has comparable development experience, so the experimental group has similar capability to the control group in total.
Note that we did not ask participants to finish half of the tasks with our tool while the other half without assisting tool to avoid potential tool bias.
We recorded the time used to reproduce the bug recordings in Android.
Participants had up to 10 minutes for each bug replay.
To minimize the impact of stress, we gave a few minutes break between each bug replay.
At the end of the tasks, we provided 5-point Likert-scale questions to collect their feedback, in terms of clearness, conciseness, and usefulness. 
We further collected participants’ feedback through a few open-ended questions, which can help us bring more insight into our tool, including how could the subtitles be improved, are there any software engineering tasks that would benefit from subtitles, etc.

\begin{figure}  
	\centering 
	\includegraphics[width=0.75\linewidth]{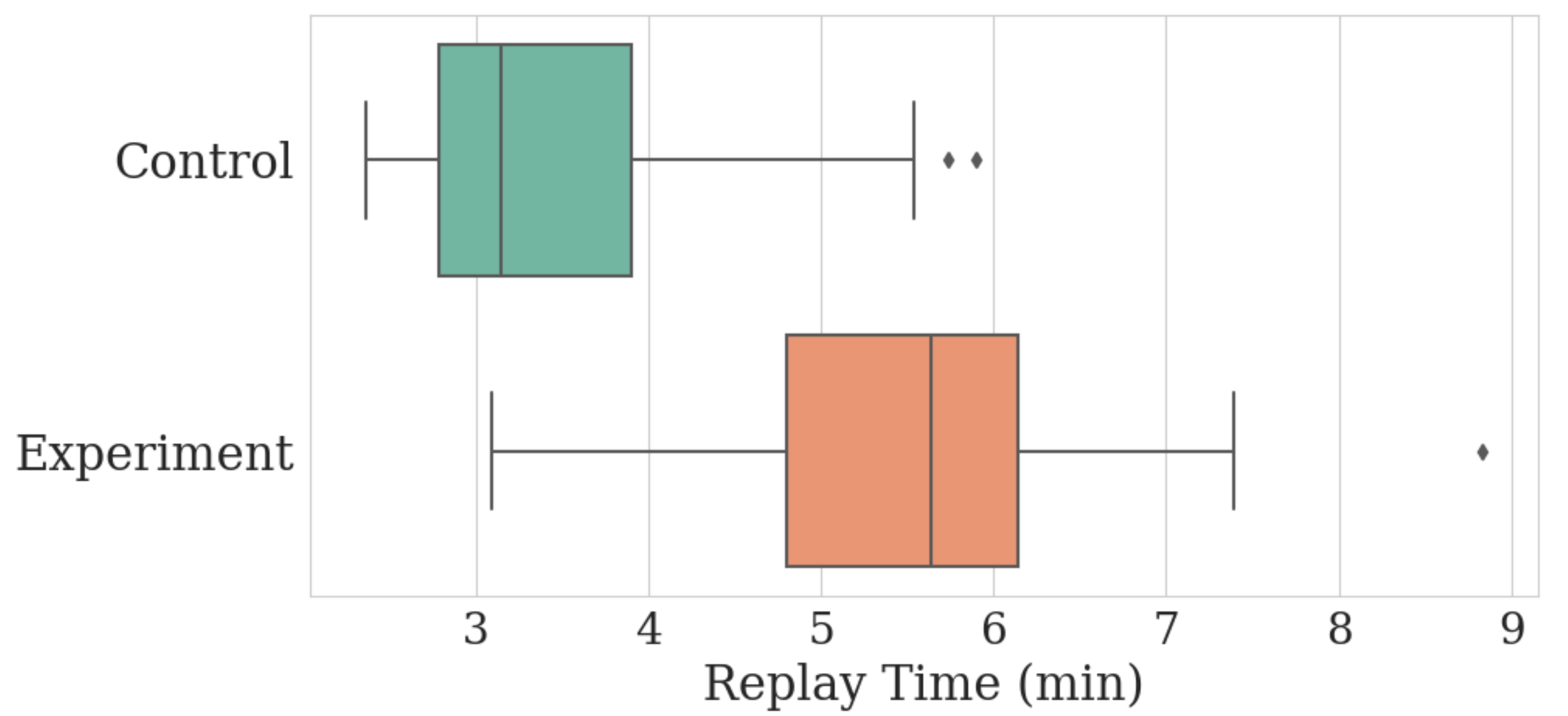}
	\caption{Bug replay time.} 
	\label{fig:reptime}
\end{figure}


\renewcommand{\arraystretch}{1}
\begin{table}
	\centering
	\caption{Performance comparison between the experimental and control group. $^*$ denotes \textit{p} $<$ 0.01.}
	\label{tab:user_study}
	\begin{tabular}{l|c|c} 
        \hline
        \bf{Measures} & \bf{Control} & \bf{Experiment} \\ 
        \hline
        Clearness & 2.50 & 4.25$^*$ \\
        Conciseness & 1.75 & 4.50$^*$ \\
        Usefulness & - & 4.75 \\
        \hline
        \end{tabular}
\end{table}

\textbf{Results:}
Overall, participants appreciate the usefulness of our approach for providing them with clear and concise step descriptions to describe the actions performed on the bug recordings, so that they can easily replay them.
Box plot in Fig.~\ref{fig:reptime} shows that, given our generated reproduction steps, the experimental group reproduces the bug recording much faster than that of the control group (with an average of 3.46min versus 5.53min, saving 59.8\% of the time).
This brings a preliminary insight of the usefulness of our generated reproduction steps to help participants locate and replay the actions.

Table~\ref{tab:user_study} shows the overall results received from participants.
All participants admit that our approach can provide more easy-to-understand step descriptions for them, in terms of 4.25 vs 2.50 in clearness, and 4.50 vs 1.75 in conciseness, compared with the control group.
In addition, they demonstrate several advantages of our reproduction steps, such as complete steps, region/text of interest, technical language, etc.
Since the steps we generate are matched with each action one-to-one, participants can easily track each step, while the missing steps in the control group may confound participants: whether the step description corresponds to the current GUI.
$P_5$ also finds the absolute positioning and element relationship description particularly useful to him, because such description can narrow down the spatial regions in GUI and easily locate the GUI element in which a bug occurs.
$P_3$ reports that some users may use inconsistent words to describe the steps.
For example, users may use ``play the film'' to describe the button with the text ``movie'', making the developers hard to reproduce in practice.
In contrast, the descriptions we generate are entirely based on GUI content, so it is easy to find the GUI elements.

The participants strongly agree (4.75) with the usefulness of our approach due to two reasons.
One is the potential of our structured text to benefit short- and long-term downstream tasks, such as bug triaging, test migration, etc.
The potential downstream is discussed in Section~\ref{sec:discussion}.
The other is the usefulness of the subtitle in the recording, revealing the action segmentation of our approach.
$P_2$ in the control group finds the touch indicator to be inconspicuous and sometimes GUI transitions are too abrupt to realize. 
In contrast, with the help of our approach, $P_6$ praises the subtitle in the recording as it informs the timing of each action.

To understand the significance of the differences, we further carry out the Mann-Whitney U test~\cite{fay2010wilcoxon} (specifically designed for small samples) on the replaying time, clearness, conciseness, and usefulness between the experimental and the control group respectively.
The test results suggest that our approach does significantly help the participants reproduce bug recordings more efficiently ($p < 0.01$).
There is also some valuable feedback provided by the participants to help improve the \tool.
For example, participants want higher-level semantic step descriptions, e.g., tap the first item in the list group, which can lead to more insights into the bugs.
We will investigate the possible solution as our future work.

\section{Discussion}
\label{sec:discussion}
We have discussed the limitations of our approach at the end of each subsection of the evaluation in Section~\ref{sec:evaluation}, such as errors due to slow rendering in action segmentation (Section~\ref{sec:evaluation_1}), low contrast between touch indicators and icons in action attribute inference (Section~\ref{sec:evaluation_2}), etc.
In this section, we discuss the implication of our approach and future work.

\textbf{Downstream tasks supported by video captioning.}
There are many downstream tasks based on the textual bug reports, such as automated bug replay~\cite{zhao2019recdroid,fazzini2018automatically}, test migration~\cite{talebipour2021ui,zhao2020fruiter}, duplicate bug detection~\cite{hindle2019preventing,xie2020uied,nguyen2012duplicate}, etc. Few of them can be applied to visual bug recordings.
Our approach to automatically caption bug recording provides a semantic bridge between textual and visual bug reports.
In detail, \tool complement the existing methods, as the first process of these downstream tasks is usually to employ natural language processing (NLP) techniques to extract the representations of bug steps into a structural grammar, such as action, object, and position, which can be automatically extracted by our approach in visual bug recording.

\textbf{Generality across platforms.}
Results in the usefulness evaluation in Section~\ref{sec:user_study} have demonstrated the usefulness of our approach in generating high-quality descriptions for Android bug recordings to help developers with bug replay in real-world practice.
Supporting bug recordings of different platforms (e.g., iOS, Web) can bring analogous benefits to developers~\cite{cooper2021takes}.
As the actions from different platforms exert almost no difference, and our approach is purely image-based and non-intrusive, it can be generalized to caption bug recordings for other platforms with reasonable customization efforts to our approach.
In the future, we will conduct thorough experiments to evaluate the performance of \tool in supporting those platforms.

\textbf{Accessibility of GUI recording.}
Tutorial videos (e.g., app usage recordings) are widely used to guide users to access unfamiliar functionalities in mobile apps.
However, it is hard for people with vision impairments (e.g., the aged or blind) to understand those videos unless asking for caregivers to describe the action steps from the tutorial videos to help them access the video content~\cite{liu2021makes}.
Our approach might be applied to enhance the accessibility of tutorial videos by generating clear and concise subtitles for reproduction steps, enabling people with vision impairments to easily access information and service of the mobile apps for convenience.

\section{Related Work}
Vision to Language semantically bridges the gap between visual information and textual information.
The most well-known task is image captioning, describing the content of an image in words.
Many of the studies proposed novel methods to generate a textual description for GUI image, in order to enhance app accessibility~\cite{chen2020unblind,li2020widget,feng2021auto,feng2022auto}, screen navigation~\cite{li2019pumice,li2018kite}, GUI design search~\cite{chen2019gallery,feng2022gallery,chen2020lost}, automate testing~\cite{liu2022fill,xie2022psychologically,liu2022guided,liu2020owl,su2022metamorphosis,liu2022nighthawk}, etc.
Chen et al.~\cite{chen2018ui} designed an approach that uses a machine translator to translate a GUI screenshot into a GUI skeleton, a functional natural language description of GUI structure.
Moran et al.~\cite{moran2022empirical} proposed image captioning methods Clarity to describe the GUI functionalities in varying granularity.
In contrast, we focused on a more difficult task - video captioning, generating natural language to describe the semantic content of a sequence of images. 
To the best of our knowledge, this is the first work translating the GUI recording into textual descriptions.

Earlier works~\cite{ma2002user,ma2005generic} proposed sequence-to-sequence video captioning models that extract a sequence of image features to generate a sequence of text. 
These models showed their advantage in video summarization, but it was hard to achieve the goal of generating multiple concrete captions with their temporal locations from the video (a.k.a dense video captioning).
Intuitively, dense video captioning can be decomposed into two phases: event segmentation and event description.
Existing methods tackled these two sub-problems using event proposal and captioning modules, and exploited two ways to combine them for dense video captioning.
We borrowed the two-phase idea to generate a natural language description for GUI recording, denoting events as user actions.

To segment the events from the videos, Krishna et al.~\cite{krishna2017dense} proposed the first segmentation method by using a multi-scale proposal module.
Some of the following works~\cite{wang2018bidirectional,sun2019videobert} aimed to enrich the event representations by context modeling, event-level relationships, or multi-modal feature fusion, enabling more accurate event segmentation.
However, these methods were designed for general videos which contain more natural scenes like human, plants, animals, etc. Different from those videos, our GUI recordings belonged to artificial artifacts with different image motions (i.e., GUI rendering). 
While some previous studies worked on domain-specific GUI recordings, they focused on high-level GUI understanding, such as duplicate bug detection~\cite{cooper2021takes}, GUI animation linting~\cite{zhao2020seenomaly,feng2022efficiency}, etc.
In contrast, we focused on the  fine-grained user actions in the GUI recording.
To analyse and segment actions from the GUI recording, many record-and-replay tools were developed based on different types of information, including the runtime information~\cite{gomez2013reran} and app artifacts~\cite{krieter2018analyzing, feng2022gifdroid,feng2022gifdroid2}.
Nurmuradov et al.~\cite{nurmuradov2017caret} introduced an advanced lightweight tool to record user interactions by displaying the device screen in a web browser. 
Feng et al.~\cite{feng2022gifdroid,feng2022gifdroid2} proposed an image processing method to extract the keyframes from the recording and mapped them to states in the GUI transitions graph to replay the execution trace.
However, they required the installation of underlying frameworks, or instrumenting apps which is too heavy and time-consuming.
Bernal et al.~\cite{bernal2020translating} implemented a deep learning-based tool named V2S to detect and classify user actions from specific recordings, a high-resolution recording with a default Android touch indicator.
But more than 32\% of end-users cannot meet that requirement in real-world bug reports according to our analysis in Section~\ref{sec:background}.
In contrast, considering the diversity of touch indicators in the general GUI recordings from end-users, we propose a more advanced approach to capture the spatial features of touch indicators and the temporal features of touch effects, to achieve better performance on user action identification.


To generate video captions, many works~\cite{ sun2019videobert,zhu2020actbert} started using one single unified deep-learning model (one-fit-all).
Recent works infused knowledge about objects in the video by using object detectors to generate more informative captions.
For example, Zhang et al.~\cite{zhang2020object} adopted an object detector to augment the object feature to yield object-specific video captioning.
Different from the natural scenes, generating action-centric descriptions for GUI recording requires a more complex GUI understanding, as there are many aspects to consider, such as the elements in the GUI, their relationships, the semantics of icons, etc.
Therefore, we modeled GUI-specific features by using mature methods, and then proposed a tailored algorithm to automatically generate natural language descriptions for GUI recordings.
\section{Conclusion}
The bug recording is trending in bug reports due to its easy creation and rich information. 
However, watching the bug recordings and understanding the user actions can be time-consuming.
In this paper, we present a lightweight approach \tool to automatically generate semantic descriptions of user actions in the recordings, without requiring additional app instructions, recording tools, or restrictive video requirements.
Our approach proposes image-processing and deep-learning models to segment bug recordings, infer user actions, and generate natural language descriptions.
The automated evaluation and user study demonstrate the accuracy and usefulness of \tool in boosting developers’ productivity.

In the future, we will keep improving our method for better performance in terms of action segmentation and action attribute inference. 
According to user feedback, we will also improve the understanding of GUI to achieve higher-level semantic descriptions.

\bibliographystyle{IEEEtran}
	\bibliography{main}
\end{document}